\documentclass[pre,twocolumn,showpacs,superscriptaddress,preprintnumbers,floatfix]{revtex4}

\usepackage{etex}
\usepackage{ifpdf}
\usepackage{hyperref}
\usepackage{dcolumn}
\usepackage{url}
\usepackage{amsmath}
\usepackage{amscd}
\usepackage{amsfonts}
\usepackage{amssymb}
\usepackage{bm}   
\usepackage{bbm}
\usepackage{verbatim}
\usepackage{stmaryrd}
\usepackage{amsthm}
\usepackage{xcolor}
\usepackage{setspace}
\usepackage{multirow}

\ifpdf
  \usepackage[pdftex]{graphicx}
\else
  \usepackage[dvips]{graphicx}
\fi

\theoremstyle{plain}    
\theoremstyle{plain}    
\theoremstyle{plain}    
\theoremstyle{plain}    
\theoremstyle{plain}    
\theoremstyle{plain}    
\theoremstyle{plain}    \newtheorem{Prop}{Proposition}
\theoremstyle{plain}    \newtheorem*{ProProp}{Proof}
\theoremstyle{plain}    
\theoremstyle{plain}    
\theoremstyle{plain}    \newtheorem{Def}{Definition}
\theoremstyle{plain}    

\newcommand{\eM}     {$\epsilon$\protect\nobreakdash-machine}
\newcommand{\eMs}    {$\epsilon$\protect\nobreakdash-machines}
\newcommand{\EM}     {$\epsilon$\protect\nobreakdash-Machine}
\newcommand{\EMs}    {$\epsilon$\protect\nobreakdash-Machines}

\newcommand{\eTs}    {$\epsilon$\protect\nobreakdash-transducers}

\newcommand{\MeasAlphabet}	{\mathcal{A}}

\newcommand{\CausalState}	{ \mathcal{S} }

\newcommand{\causalstate}	{ \sigma }
\newcommand{\CausalStateSet}	{ \boldsymbol{\CausalState} }

\newcommand{\Prob}		{ {\Pr}}

\newcommand{\T}{\mathbb{T}}
\newcommand{\M}{\mathbb{M}}

\newcommand{\PrEMType}{\mathbf{f}}

\newif\ifcolor

\colortrue

\ifcolor\relax\else\selectcolormodel{gray}\fi

\begin{document}

\title{How the Dimension of Space Affects the Products of Pre-Biotic Evolution:\\
The Spatial Population Dynamics of Structural Complexity and\\
The Emergence of Membranes}

\author{Steve T. Piantadosi}
\email{piantado@mit.edu}
\affiliation{Department of Brain and Cognitive Science,\\
Massachusetts Institute of Technology, 43 Vassar Street, Cambridge, MA 02139}
\affiliation{Santa Fe Institute, 1399 Hyde Park Road, Santa Fe, NM 87501}

\author{James P. Crutchfield}
\email{chaos@ucdavis.edu}
\affiliation{Santa Fe Institute, 1399 Hyde Park Road, Santa Fe, NM 87501}
\affiliation{Complexity Sciences Center and Physics Department,\\
University of California at Davis, One Shields Avenue, Davis, CA 95616}

\date{\today}

\bibliographystyle{unsrt}

\begin{abstract}
We show that autocatalytic networks of \eMs\ and their population dynamics
differ substantially between spatial (geographically distributed) and nonspatial
(panmixia) populations. Generally, regions of spacetime-invariant autocatalytic
networks---or \emph{domains}---emerge in geographically distributed populations.
These are separated by \emph{functional membranes} of complementary \eMs\ that
actively translate between the domains and are responsible for their growth and
stability. We analyze both spatial and nonspatial populations, determining the
algebraic properties of the autocatalytic networks that allow for space to
affect the dynamics and so generate autocatalytic domains and membranes. In
addition, we analyze populations of intermediate spatial architecture,
delineating the thresholds at which spatial memory (information storage)
begins to determine the character of the emergent auto-catalytic organization.

\vspace{0.1in}
\noindent
{\bf Keywords}: population dynamics, structural complexity, spatial dimension,
autocatalytic network

\end{abstract}

\pacs{
87.18.-h  
87.23.Cc  
87.23.Kg  
05.45.-a  
89.75.Kd  
}
\preprint{Santa Fe Institute Working Paper 10-09-XXX}
\preprint{arxiv.org:1009.XXXX [q-bio-mat.XXXX]}

\maketitle



%

\section{Introduction}

Almost certainly, the sophisticated mechanisms for self-replication found
today in living cells were not present in the earliest replicators
\cite{Szathmary87,Mayn95a,Gest99a}. Instead, some hypothesized that the first
replicators were autocatalytic networks of functional molecules that
collectively were capable of self-reproduction \cite{Eigen71,Eigen77}. Numerous
studies have been devoted to these and analogous networks with the hope of
understanding pre-biotic evolution \cite{Ray91a,Adami1,Rasmussen2}. A particular
class of such models represented network elements with mathematical constructs,
including regular expressions, the $\lambda$-calculus \cite{Font91a}, and,
recently, \eMs\ \cite{OTMOMerge}. \EMs\ are especially useful
since they support well defined and computable measures of structural complexity
\cite{Crut08b} and, equally important, these measures extend directly to
networks of interacting \eMs. 

Early models often assumed a \emph{Turing gas} in which every network element
has an equal chance of interacting with every other, thereby ignoring a
network's spatial configuration. However, natural and engineered evolutionary
systems are not architected this way since elements of physical systems always
have some spatial relationship that determines which components interact.
This observation led to studies of spatial pattern formation in evolutionary and
autocatalytic systems \cite{Kirn00a,Kirn02a,Acke06a}.

Here, we consider networks of single-state \eMs---a simplifying initial focus
that helps to highlight the role of population architecture. We show that the
behavior of spatially distributed populations differs substantially from that
of the nonspatial Turing gas. More to the point, we determine the algebraic
properties of the network that lead to the emergence of distinctive
organizations.

\vspace{-0.2in}
\section{Background}
\vspace{-0.1in}

\emph{\EMs} serve as models of stochastic finite and infinite computation
\cite{CMechMerge}. An \eM\ consists of a set $\CausalStateSet$ of
\emph{causal states} and a set of \emph{transitions} between those states. In
the particular variation used here (\eTs), each transition is labeled by both an
input symbol $x \in \MeasAlphabet$ and an output symbol $y \in \MeasAlphabet$.
An \eM\ in state $\causalstate \in \CausalStateSet$ reads an input symbol $x$
and chooses from all transitions from $\causalstate$ the one labeled $x$.
The \eM\ follows the chosen transition to a next state
$\causalstate^\prime \in \CausalStateSet$ while emitting the
output symbol $y$ corresponding to that transition. In this way, \eMs\ can be
viewed as mapping an input language to an output language, perhaps
probabilistically. 

Following Ref. \cite{OTMOMerge}, we focus on single-state
\eMs\ over the input-output alphabet $\MeasAlphabet = \lbrace 0, 1 \rbrace$.
(Results for multi-state \eMs\ appear in a sequel. Our goal here is to
highlight the effects of space, uncomplicated by the richness that comes with
using multi-state \eMs.) There are $16$ such \eMs\ and we denote the set of all
them by $\T$. Each \eM\ can be represented as a $2 \times 2$ binary matrix $M$,
where $M_{ij} = 1$ means that the \eM\ reads in symbol $i-1$ while emitting
symbol $j-1$. We number the $16$
\eMs---$\T \equiv \{ T_0, \ldots, T_{15} \}$---by
finding the decimal equivalent of the binary number
$M_{11} M_{12} M_{21} M_{22}$ for
each \eM. Thus, for example, \eM\ $T_{11}$ has the matrix representation:
\begin{equation}
M =
  \left[
  \begin{array}{cc}
	1&0\\
	1&1
  \end{array}
  \right] ~.
\end{equation}

\EM\ pairs interact in a population by composition, under which $\T$ is closed
and forms a monoid \cite{Holc82a}. With the matrix representation,
\eM\ composition---$T_b \circ T_a$---is simply matrix multiplication where,
after multiplying, any positive matrix element is set to $1$. From this, it is
straightforward to compute the \emph{interaction matrix} $\M$, where
$T_a \circ T_b = T_c$ if and only if $\M_{a+1, b+1} = c$,
where $a,b,c \in \{0, \ldots, 15\}$:
\begin{equation}
\M =
\left[
\begin{smallmatrix}
0&0&0&0&0&0&0&0&0&0&0&0&0&0&0&0\\
 \noalign{\medskip}
0&1&2&3&0&1&2&3&0&1&2&3&0&1&2&3\\
 \noalign{\medskip}
0&0&0&0&1&1&1&1&2&2&2&2&3&3&3&3\\
 \noalign{\medskip}
0&1&2&3&1&1&3&3&2&3&2&3&3&3&3&3\\
 \noalign{\medskip}
0&4&8&12&0&4&8&12&0&4&8&12&0&4&8&12\\
 \noalign{\medskip}
0&5&10&15&0&5&10&15&0&5&10&15&0&5&10&15\\
 \noalign{\medskip}
0&4&8&12&1&5&9&13&2&6&10&14&3&7&11&15\\
 \noalign{\medskip}
0&5&10&15&1&5&11&15&2&7&10&15&3&7&11&15\\
 \noalign{\medskip}
0&0&0&0&4&4&4&4&8&8&8&8&12&12&12&12\\
 \noalign{\medskip}
0&1&2&3&4&5&6&7&8&9&10&11&12&13&14&15\\
 \noalign{\medskip}
0&0&0&0&5&5&5&5&10&10&10&10&15&15&15&15\\
 \noalign{\medskip}
0&1&2&3&5&5&7&7&10&11&10&11&15&15&15&15\\
 \noalign{\medskip}
0&4&8&12&4&4&12&12&8&12&8&12&12&12&12&12\\
 \noalign{\medskip}
0&5&10&15&4&5&14&15&8&13&10&15&12&13&14&15\\
 \noalign{\medskip}
0&4&8&12&5&5&13&13&10&14&10&14&15&15&15&15\\
 \noalign{\medskip}
0&5&10&15&5&5&15&15&10&15&10&15&15&15&15&15\\
 \noalign{\medskip}
\end{smallmatrix}
    \right] ~.
\end{equation}

\section{\EM\ Soups}

A population, or simply a \emph{soup}, $\Gamma$ is a configuration of an
$n \times n$ regular toroidal lattice. At each time $t=0,1,2,\ldots$, every
lattice location $(i,j)$ contains a single \eM, denoted $\Gamma_{i,j}^t \in \T$.
The population size is $N = n^2$. Each location $(i,j)$ is initialized to
contain an \eM\ uniformly chosen at random from $\T$.

The population dynamics is specified by \eM\ composition, interaction, and
update. We define a function $\theta: \T \times \T \times \T \to \T$ by:
\begin{equation}
\theta(T_{a}, T_{b}, T_{c}) =
  \begin{cases}
    T_{a} \circ T_{c} & \text{ if } T_{a} \circ T_{c} \neq T_{0} \\
    T_{b} & \text{ otherwise }
  \end{cases} ~.
\end{equation}
We write $\theta(A, B, C)$ for sets $A,B,C \subseteq \T$ as shorthand for the
set $\lbrace \theta(a,b,c) : a \in A, b \in B, c \in C \rbrace$. We must use
$\theta$ to prevent the non-\eM\ $T_0$ from being produced since the fact
that $T_0 \circ T_a = T_a \circ T_0 = T_0$ for all $T_a\in \T$ implies that
if $T_0$ could be produced, it comes to dominate any population. 

Each time step $t$ we choose a location $(i,j)$ at random and set:
\begin{equation}
\Gamma_{i,j}^{t+1} =
  \begin{cases}
    \theta \left( \Gamma_{i-1,j}^{t}, \Gamma_{i,j}^t, \Gamma_{i+1,j}^{t} \right)
	& \!\!  \text{with probability } \tfrac{1}{4}\\
    \theta \left( \Gamma_{i+1,j}^{t}, \Gamma_{i,j}^t, \Gamma_{i-1,j}^{t} \right)
	& \!\!  \text{with probability } \tfrac{1}{4}\\
    \theta \left( \Gamma_{i,j-1}^{t}, \Gamma_{i,j}^t, \Gamma_{i,j+1}^{t} \right)
	& \!\!  \text{with probability } \tfrac{1}{4}\\
    \theta \left( \Gamma_{i,j+1}^{t}, \Gamma_{i,j}^t, \Gamma_{i,j-1}^{t} \right)
	& \!\!  \text{with probability } \tfrac{1}{4}\\
  \end{cases}
\label{eq:localupdate}
\end{equation}
and set $\Gamma_{k,l}^{t+1}=\Gamma_{k,l}^t$ for all $(k,l) \neq (i,j)$. Thus,
two vertical or horizontal neighbors to $\Gamma_{i,j}$ are chosen, composed,
and the \eM\ resulting from their composition is used to replace
$\Gamma_{i,j}^t$, if it is not $T_0$. This replacement scheme is meant to be
locally analogous to the replacement scheme used in Ref. \cite{OTMOMerge},
which will facilitate direct comparisons in the following.

In addition, at each time step a certain amount of diffusion of \eMs\ occurs
due to spatial mixing. For this, let $\zeta_v$ be a Gaussian distribution with
variance $v$ and mean $0$. At each time step, $c$ \eMs\ are chosen at random
in $\Gamma$ and each is swapped with a random \eM\ at a distance chosen from
the distribution $\zeta_v$. For more generality, we allow $c$ to be any real
number and swap $c$ \eMs\ per time step on average. 

When $v$ and $c$ are large, there is considerable spatial mixing and one
expects the dynamics to behave like a nonspatial population, where
\eMs\ have an equal chance of interacting with every other. However, when $v$
and $c$ are small, there is little spatial mixing and, as we shall see, the
population dynamics change substantially. 

To summarize, as a stochastic dynamical system the soup's state at time $t$ is
the population's \emph{configuration} $\Gamma^t$. While
Eq. (\ref{eq:localupdate}) determines the local probabilistic update at each
site, we use $\Theta$ to formally denote the global (probabilistic) update for
the entire configuration over one time step:
\begin{equation}
\Gamma^{t+1} = \Theta \circ \Gamma^t ~.
\end{equation}
And so, one goal is to understand the trajectories
$\{\Gamma^0, \Gamma^1, \ldots \}$. Another is to analyze the structure inside
the $\Gamma^t$. For this, we use $\sigma_{\widehat{n}}$ to denote a spatial shift
of the soup configuration:
\begin{equation}
\Gamma^\prime = \sigma_{\widehat{n}} \circ \Gamma ~,
\end{equation}
where $\widehat{n} = (\Delta i, \Delta j)$ is the vector by which the configuration
is shifted horizontally and vertically:
\begin{align}
\Gamma_{k,l}^\prime
  & = (\sigma_{\widehat{n}} \circ \Gamma)_{k,l} \nonumber \\
  & = \Gamma_{(i+\Delta i) \mathrm{~mod~} n, (j+\Delta j) \mathrm{~mod~} n} ~.
\end{align}

Finally, we view the population either as a configuration $\Gamma$ of spatially
positioned \eMs\ or as a distribution $\PrEMType$ of \eM\ types without regard
to spatial location. The fractions
$\PrEMType = \left(\PrEMType^1, \ldots, \PrEMType^{15} \right)$ of \eMs\ of
type $T_a$ on $\Gamma$ are $\PrEMType^a (\Gamma) = \Prob(T_a \in \Gamma)$.

\section{The Panmixia Soup}

The least-spatial architecture of $\Gamma$, where $c \approx N$ and
$v \approx n$, was previously studied in Ref. \cite{OTMOMerge}.
This is the case of a \emph{panmixia} population, in which all \eMs\ can
interact with any other. When the size of $\Gamma$ is large,
$\PrEMType(t) \equiv \PrEMType(\Gamma^t)$ follows a simple population dynamics.

First, for large $N$ the discrete-time dynamic is well
approximated by continuous time, since each discrete time updates
only a single lattice location and this means that changes to
$\Gamma^t$ and so to $\PrEMType(t)$ are relatively small ($\propto N^{-1})$.
Second, the probability of adding an \eM\ $T_i$ is that of picking
two \eMs\ $T_a$ and $T_b$, such that $T_a \circ T_b = T_i$, times the
probability that the \eM\ replaced is not $T_i$. Third, the probability of
removing an \eM\ $T_i$ is that of picking two \eMs, $T_a$ and $T_b$, such that
$T_a \circ T_b \neq T_i$ and also $T_a \circ T_b \neq T_0$, times the
probability of picking a $T_i$ to replace. Thus, the rate of change of
$\PrEMType^i (t)$ is given by the differential equation:
\begin{equation}
\dfrac{d\PrEMType^i}{dt}
  = \left( 1 - \PrEMType^i \right)
  \sum_{T_a \circ T_b = T_i} \PrEMType^a \PrEMType^b
  - \PrEMType^i
  \sum_{\substack{T_a \circ T_b \neq T_i \\ T_a \circ T_b \neq T_0}}
  \PrEMType^a \PrEMType^b ~,
\label{E:drhoT}
\end{equation}
where the sums run over all pairs satisfying their subscripted condition. That
is, $\sum_{T_a \circ T_b = T_i}$ runs over all ordered pairs $(T_a, T_b)$ such
that $T_a \circ T_b = T_i$. Equation (\ref{E:drhoT}) also determines the
steady-state probability distributions of \eM\ types, which are simply
solutions of:
\begin{equation}
\dfrac{d\PrEMType}{dt} = 0 ~.
\end{equation}

\begin{figure}[h]
\centering
\includegraphics[width=3.4in]{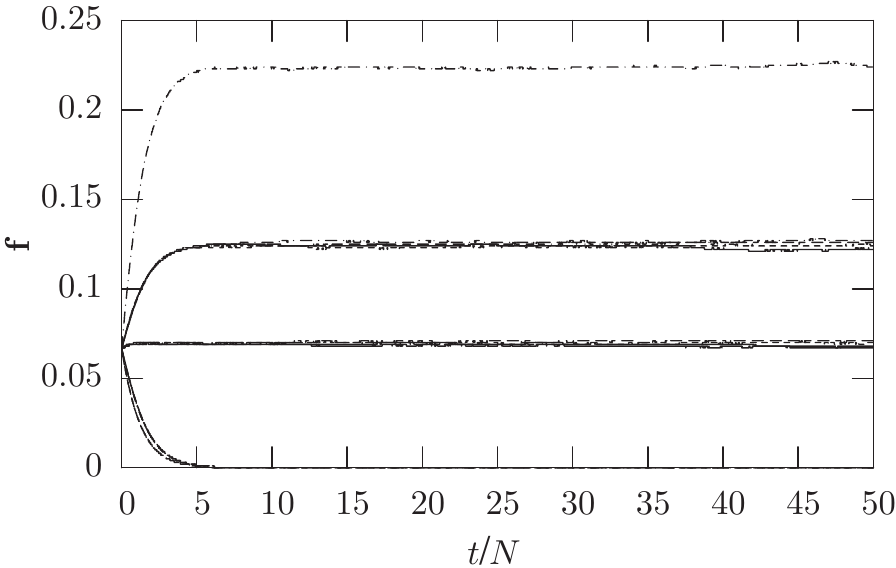}
\caption{Panmixia population evolution: \EM-type distribution $\PrEMType$
  as a function of replication time ($t/N$), with $n = 10^3$, $N = 10^6$, $c =
  3000$, and $v = 1000$.
  }
\label{WM_PD}
\end{figure}

Figure \ref{WM_PD} shows the distribution $\PrEMType (t)$ over \eM\ types as a
function of time (replication) for a simulation with $c=3000$ and $v=1000$. This
is essentially the same population behavior reported in Ref. \cite{OTMOMerge}.
It was confirmed numerically that Eq. (\ref{E:drhoT}) closely predicts the
results in Fig. \ref{WM_PD}. In Fig. \ref{WM_PD}, all $16$ single-state
\eMs\ are shown. They partition themselves into four classes such that
those in the same class behave similarly; see Table \ref{tab:PanMixiaSoup}.
Six \eMs\ die away, while nine persist in a closed, self-maintaining, and
dynamically stable \emph{meta-machine}, as shown in Ref. \cite{OTMOMerge}.

\begin{table}[tbp]
\begin{tabular}{|c||c|}
\hline
\EM\ Type                                      & Behavior Class  \\
\hline
$T_{15}$                                      & Fast Growth \\
$T_{3}, T_{5}, T_{10}, T_{12}$                & Medium Growth \\
$T_{1}, T_{2}, T_{4}, T_{8}$                  & No Growth \\
$T_{6}, T_{7}, T_{9}, T_{11}, T_{13}, T_{14}$ & Fast Decay \\
\hline
\end{tabular}
\caption{\EM-type behavior classes in the panmixia soup of Fig. \ref{WM_PD}.
  }
\label{tab:PanMixiaSoup}
\end{table}

\section{General Replicators}

While these \eM\ classes can be understood by examining the solutions of Eq.
(\ref{E:drhoT}) as done in Ref. \cite{OTMOMerge}, the results can also be
directly predicted by examining the algebraic structure of the set $\T$ under
the composition operator determined by $\M$. And this observation is critical
to predicting the behavior of alternate population architectures. First, we
address the panmixia population. The central idea is to find subsets of
\eM\ types that map onto themselves under the population dynamics; in other
words, to identify $\M$-invariant subsets of $\T$.

\begin{Def}
A set $S$ of \eMs\ is a \emph{general replicator} (GR) if for all $a \in \T$
we have:
\begin{equation*}
\theta(\T,a,S) \cup \theta(S, a, \T) \subseteq S \cup \lbrace a \rbrace
\end{equation*}
and
\begin{equation*}
\theta(\T, a, S) \cup \theta(S, a, \T) \neq \lbrace a \rbrace ~.  
\end{equation*}
\label{defDI}
\end{Def}
This parallels the definition of an \emph{ideal} in semigroup theory, except
that we must be more careful (and less elegant) with the definition since
$\theta$ is not a binary relation \cite{Holc82a}.

In $\T$ there are many GRs, including $\T$ itself. We intentionally used
$\theta$ to exclude the non-\eM\ $T_0$ from being considered a GR. By design,
$T_0$ cannot be produced. GRs are important because they are the simplest form
of replicator. Suppose $S$ is a GR. As evolution progresses, elements in the
soup are replaced with some element of $\theta(\T,a,S)$ or $\theta(S,a,\T)$.
By the definition of a GR, though, $\theta(\T,a,S)$ and $\theta(S,a,\T)$ contain
elements of $S$ and so generally we expect elements of $S$ to produce more
elements of $S$. 

However, not all GRs are equal since some are subsets of others and therefore
are produced more readily. There are, in fact, many GRs in $\T$, but there is
one GR, call it $\Omega$, that is \emph{minimal} in that no subset of $\Omega$
is a GR, but every GR contains $\Omega$. (Note that, generally, such a set is
not guaranteed to exist in a semigroup with a zero element \cite{Holc82a}.)

\begin{Prop}
$\Omega = \lbrace T_1, T_2, T_3, T_4, T_5, T_8, T_{10}, T_{12}, T_{15} \rbrace$
is the minimal general replicator in $\T$.
\end{Prop}

\begin{ProProp}
By direct verification of Def. \ref{defDI} and observing that $\Omega$, short
any one of its \eMs, is not a GR.
\end{ProProp}

In the panmixia population, we expect that elements of $\Omega$ come to
dominate the soup. And this is what is observed in the simulations
(Fig. \ref{WM_PD} and Table \ref{tab:PanMixiaSoup}). While Eq. (\ref{E:drhoT})
explains the specific values of $\PrEMType$, an understanding of $\M$'s
structure leads to a direct explanation for why the set Fast Decay is removed
from the soup: None of the \eMs\ in Fast Decay are in $\Omega$. 

\vspace{-0.2in}
\section{The Spatial Soup}
\vspace{-0.1in}

The population that we consider next is that of a spatially configured soup
where $c=0$ or $v=0$. In these limits, no spatial mixing is added to the system
and the model behaves essentially as an asynchronous, probabilistic cellular
automaton. Specifically, each element is replaced by the composition of two of
its neighbors if that composition does not result in $T_0$. The pair of
neighbors composed is chosen with uniform probability, according to
Eq. (\ref{eq:localupdate}).

\begin{figure}[hbt]
\includegraphics[width=3.4in]{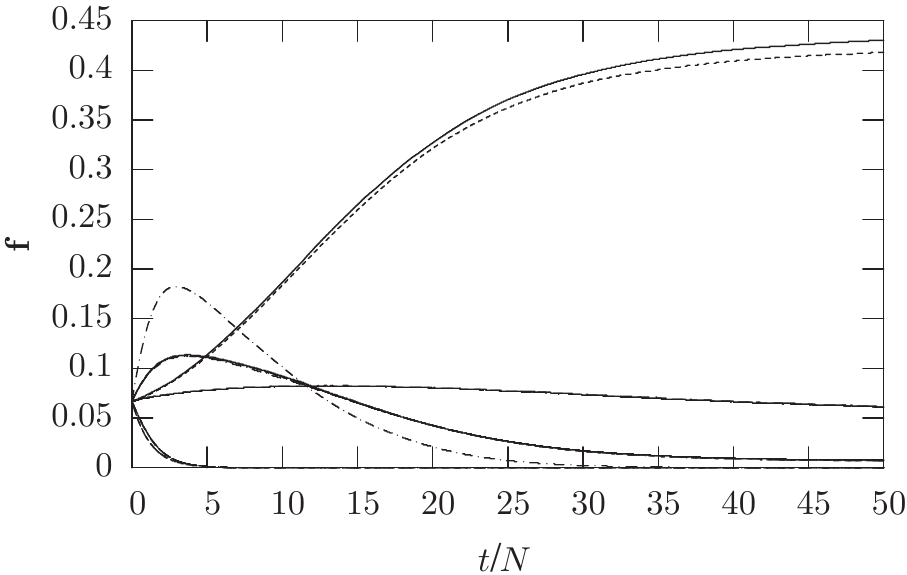}
\caption{Spatial population dynamics: \EM-type distribution $\PrEMType$ as a
  function of replication time with $n = 10^3$, $N = 10^6$, and $c=0$.
  }
\label{UM_PD}
\end{figure}

Figure \ref{UM_PD} shows the \eM-type distribution $\PrEMType (t)$ for $c=0$
and $N = 10^6$.
Note that initially the population behaves quite similarly to the panmixia
case. This is to be expected since the simulations begin with identical
initial configurations so at $t=0$ the range of possible interactions is
effectively the same for both architectures. As in the panmixia population,
elements not in $\Omega$ are quickly removed from the soup. 

\begin{table}[h]
\begin{tabular}{|c||c|c|}
\hline
\EM\ Type & Early & Late \\
\hline
$T_{15}$                                      & Fast Growth & Decay \\
$T_{3}, T_{5}, T_{10}, T_{12}$                & Medium Growth & Decay \\
$T_{2}, T_{4}$                                & Logistic Growth & Saturation \\
$T_{1}, T_{8}$                                & No Growth & No Growth \\
$T_{6}, T_{7}, T_{9}, T_{11}, T_{13}, T_{14}$ & Fast Decay & Removed \\
\hline
\end{tabular}
\caption{\EM-type behavior classes in the two-dimensional spatial population
  for early and late times for Fig. \ref{UM_PD}.
  Early times: $t/N < 3.4$; late times: $t/N > 3.4$.
  }
\label{tab:Spatial2DSoup}
\end{table}

By $t / N \approx 3.4$, however, the populations start to behave differently.
For example, the \eM\ $T_{15}$ that is most readily made in the panmixia case
is also readily made in the spatial case, until $t/N=3.4$, when it begins to
be removed from the population. Again, the \eMs\ partition themselves into
subsets whose elements behave similarly. At early times before
$t/N=3.4$, these are identical to those seen
in the panmixia population except that $T_2$ and $T_4$ undergo logistic growth,
while $T_1$ and $T_8$ do not. Table \ref{tab:Spatial2DSoup} summarizes the
overall behavior, for both early and late times.

\begin{figure*}[hbtp]
  \vspace{9pt}
  \centerline{\hbox{ \hspace{0.0in}
	\includegraphics[width=2.0in]{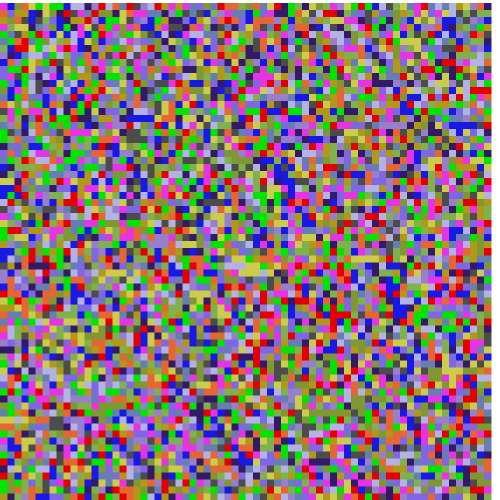}
    \hspace{10pt}
	\includegraphics[width=2.0in]{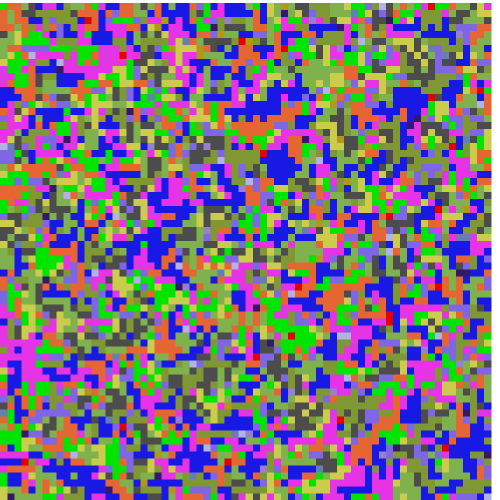}
    }
  }
  \vspace{3pt}
  \hbox{\hspace{2.1in} $t/N=0.1$ \hspace{1.5in} $t/N=3.4$}
  \vspace{15pt}
  \centerline{\hbox{ \hspace{0.0in}
	\includegraphics[width=2.0in]{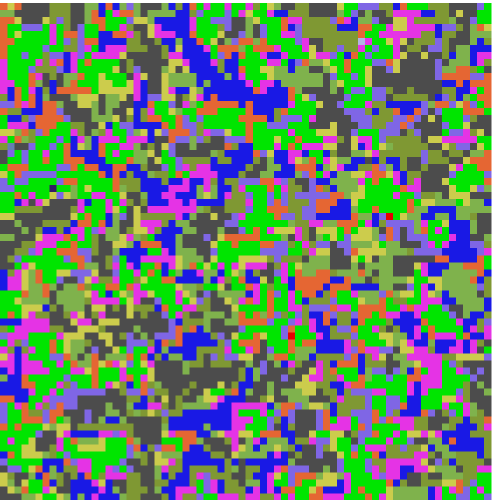}
    \hspace{10pt}
	\includegraphics[width=2.0in]{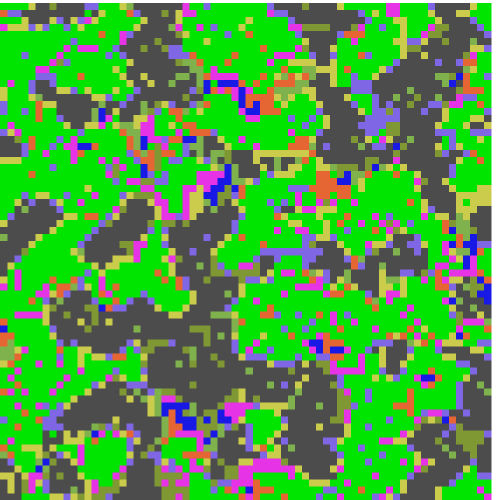}
    \hspace{10pt}
	\includegraphics[width=2.0in]{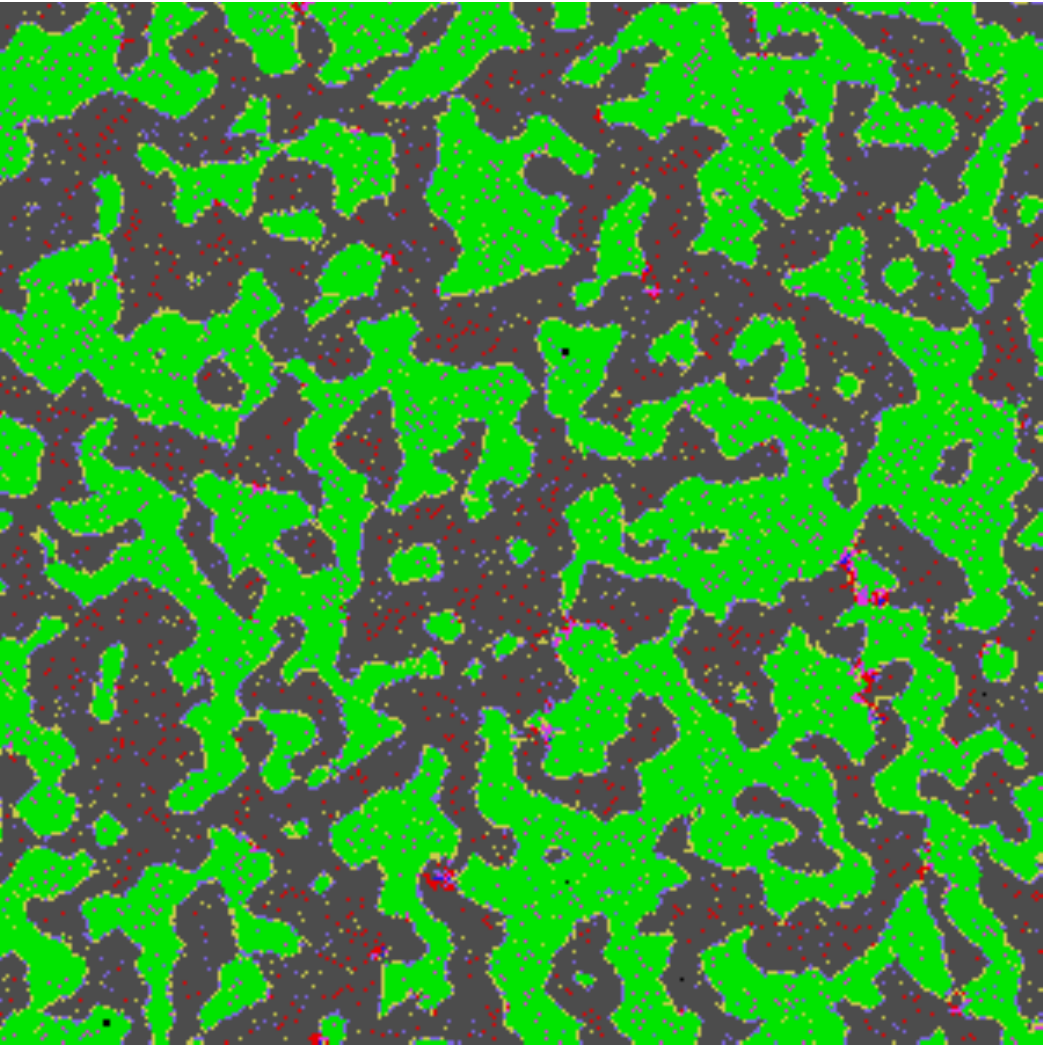}
    }
  }
  \vspace{3pt}
  \hbox{\hspace{1.0in} $t/N=9.0$ \hspace{1.2in} $t/N=20.0: 70 \times 70$
  \hspace{0.7in} $t/N=20.0: 500 \times 500$}
  \vspace{7pt}
\caption{Emergent spatial replicators and their membranes: A $70 \times 70$
  region of $\Gamma$ with $n = 10^3$, $N=10^6$, and $c = v = 0$.
  The last image, however, is of a $500 \times 500$ region.
  Each color corresponds to one of the $16$ \eM\ types: $T_2$ is green,
  $T_4$ is dark gray, $T_1$ is yellow, $T_8$ is purple, and $T_{15}$
  is blue.
  }
\label{fig:SpatialPix}
\end{figure*}

More striking than the evolution of the \eM-type distribution $\PrEMType(t)$,
however, are the spatial patterns that emerge in $\Gamma^t$. Figure
\ref{fig:SpatialPix} shows a $70 \times 70$ region of $\Gamma^t$ at increasing
times and one $500 \times 500$ region at a late time. Regions of $T_2$ and
$T_4$ form stable domains that grow to dominate the soup. Within a region of
$T_2$ or $T_4$ there exist \eMs\ that cannot be replaced, typically due to
an ``elastic'' collision---an interaction producing the non-\eM\ $T_0$
that, by $\theta$, simply leaves the \eM\ at the lattice location alone.

When a region of $T_2$ meets a region of $T_4$, \eMs\ $T_1$ and $T_8$ form on
the boundary, since $T_2 \circ T_4 = T_1$ and $T_4 \circ T_2 = T_8$. Moreover,
$T_1 \circ T_1 = T_1$ and $T_8 \circ T_8 = T_8$ and so the boundaries are
self-sustaining along the interface.

The spatial soup exhibits many of the nontrivial spontaneous patternings
common to reaction-diffusion systems that exhibit Turing instability
\cite{Turi52,Mein82a,Ball99a,Hoyl06a}. Finding a set of reaction-diffusion
partial differential equations equivalent to this model and the spatial analog
of Eq. ~\ref{E:drhoT}, however, remains an open problem. 

Over a large number of time steps, the spatial population looks essentially
like that at $t/N=20.0$ in Fig. \ref{fig:SpatialPix}. If enough time passes,
though, the soup eventually divides itself into two connected regions of $T_2$
and of $T_4$, or one will take over completely. This requires an extremely
large number of replications. Which \eM\ region dominates in the long run
appears to be randomly determined. The overall process is highly reminiscent of
spinodal decomposition in which a mixed solution separates into stable component
phases \cite{Favv08a}. A more direct connection to the predictions of that
theory awaits further effort.

\vspace{-0.2in}
\section{Spatial-Replicator Domains and Their Membranes}
\vspace{-0.1in}

After time $t/N \approx 3.4$, the characters of the spatial and panmixia
populations begin to diverge substantially, with patterns emerging in the
spatial $\Gamma$. Those patterns consist of \emph{domains} of nearly homogeneous
\eMs\ of type $T_2$ or of type $T_4$. They are separated by domain \emph{walls}
that consist of \eMs\ $T_1$ and $T_8$. As we shall see, these walls play the
role of functional membranes that actively translate between the domain \eMs.

\begin{figure*}[hbtp]
  \vspace{9pt}
  \centerline{\hbox{ \hspace{0.0in}
	\includegraphics[width=2.0in]{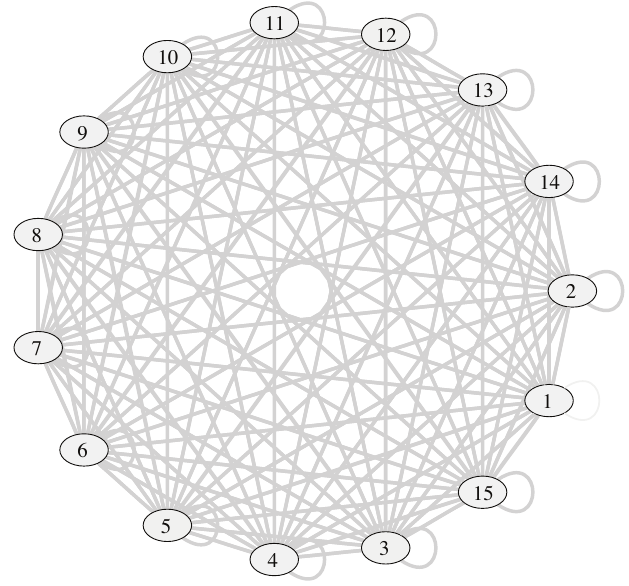}
    \hspace{10pt}
	\includegraphics[width=2.0in]{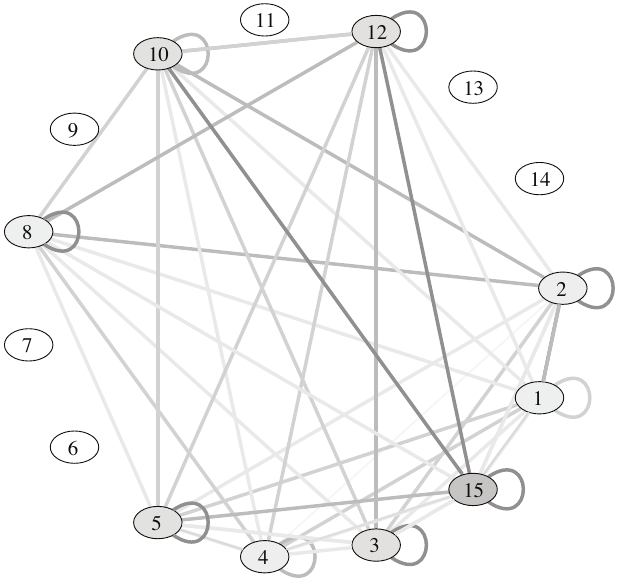}
    }
  }
  \vspace{3pt}
  \hbox{\hspace{2.2in} $t/N = 0$ \hspace{1.8in} $t/N = 5$}
  \vspace{15pt}
  \centerline{\hbox{ \hspace{0.0in}
	\includegraphics[width=2.0in]{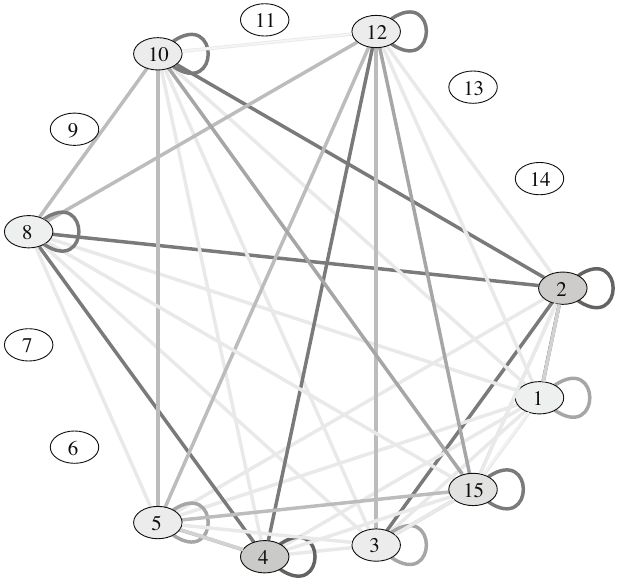}
    \hspace{10pt}
	\includegraphics[width=2.0in]{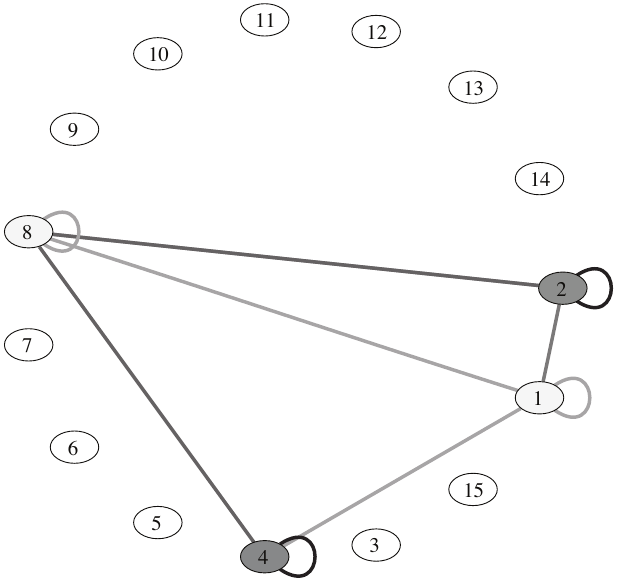}
    }
  }
  \vspace{3pt}
  \hbox{\hspace{2.1in} $t/N = 25$ \hspace{1.8in} $t/N = 75$}
  \vspace{7pt}
\caption{The effective \eM\ interactions at various times.
  The nodes are the $15$ possible \eMs. The darkness of their
  interiors indicates their relative fractions $\PrEMType^i$ in the
  populations. The edges connecting them indicate the frequency of
  their interactions; darker is more frequent.
  }
\label{fig:InteractionVTime}
\end{figure*}

To summarize, then, the first observation about the spatial population is that
it is completely differently organized. The persistent set of \eMs\ differs
markedly from those found with panmixia, as do their roles and interactions.
Why did they emerge when the population is embedded in space? We can explain
the emergence of these structures by defining a mechanism in the interaction
network $\M$ that affects the population dynamics when there is sufficient
``spatial memory''. This mechanism is called a \emph{spatial replicator} (SR).

\begin{Def}
A set $S$ of \eMs\ is a \emph{spatial replicator} (SR) if 
\begin{align*}
\theta(S,S,\theta(S,S,\T)) & \subseteq S,\\
\theta(S,S,\theta(\T,S,S)) & \subseteq S,\\
\theta(\theta(S,S,\T),S,S) & \subseteq S, \mathrm{ and} \\
\theta(\theta(\T,S,S),S,S) & \subseteq S.
\end{align*}
\end{Def}

Notice that this is similar to the definition of a GR, except that SRs require
two applications of $\theta$ to produce an element of $S$. Due to this, SRs
only replicate when there is spatial information storage: If $S$ is a SR, an
element of $S$ must compose with an \eM\ in $\T$ to produce an \eM\ $T_y$.
$T_y$ must then compose with an element of $S$ again to produce a new element
of $S$. 

In the panmixia population there is no notion of adjacency. Therefore, $T_y$
does not have an increased chance of interacting with an element of $S$ again.
However, space allows a way for SRs to replicate: If by chance a sufficiently
large domain of $\Gamma$ consists only of elements of some SR $S$, \eMs\ on
the domain border will be replaced by elements of $\theta(S, S, \T)$ or
$\theta(\T, S, S)$. Later, these boundary \eMs\ will be replaced with elements of 
\begin{align}
\theta(S,S, & \theta(S,S,\T)) \cup \theta(S,S,\theta(\T,S,S))
  \nonumber \\
  & \cup \theta(\theta(S,S,\T),S,S) \cup \theta(\theta(\T,S,S),S,S) ~, 
\end{align}
which are all elements of $S$. Thus, a domain of $S$ will grow. This argument
lays out how the algebraic structure that space adds to $\M$ leads to domain
growth. In addition, since $\theta(S, S, S) = S$, domains of $S$ are
self-maintaining.

As with a GR, one expects that if $S'$ is a SR and is also a subset of some
other SR $S$, then $S'$ eventually will come to dominate the soup. For the
spatial population of single-state \eMs, it is easy to check that
$\{ T_1, T_2, T_4, T_8 \}$ is a SR. Dynamically, the membranes of $T_1$ and
$T_8$ grow their respective domains of $T_4$ and $T_2$. Unsurprisingly, then,
$T_2$ and $T_4$ come to dominate $\Gamma$. In contrast, $T_{15}$---an \eM\ made
so readily by the panmixia population---begins to dominate the simulation until
``spatial memory'' begins to act. Beyond that point $T_{15}$ is replaced by SRs.

\begin{figure}[hbtp]
  \includegraphics[width=2.5in]{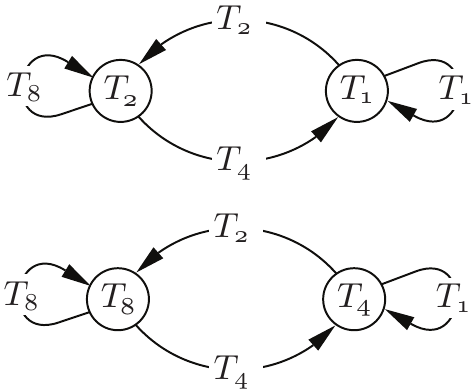}
\caption{The \eM\ interaction network at long times in the spatial soup. An
  arrow labeled $T_b$ going from node $T_a$ to node $T_c$ denotes
  $T_c = T_b \circ T_a$.
  }
\label{fig:SpatialMetaMachine}
\end{figure}

Figure \ref{fig:InteractionVTime} shows how these interactions develop over
time by giving the effective \eM\ interactions on an undirected graph, where
the darkness of the \eM\ nodes indicates their relative frequencies
$\PrEMType^i$ and the darkness of the connecting edges indicates the relative
frequency of interactions. As a complement to this,
Fig. \ref{fig:SpatialMetaMachine} gives
the \eM\ interaction network at long times. This should be compared to the
nine-\eM\ meta-machine in the panmixia soup: see Fig. 3 of Ref. \cite{OTMOMerge}.

The explanation for why a spatial population organizes differently
and, in effect, selects different individuals for different roles
relies on SRs, a local mechanism. The domains and walls, however,
are nonlocal structures. Given a local dynamic that produces them,
how do we describe their global structure?

To do this we adapt the computational mechanics analysis of emergent domains and
particles in deterministic, synchronous cellular automata (CAs)
\cite{CACMechMerge} to the stochastic, asynchronous spatial
population dynamics here. Speaking informally, a domain is a patch of spacetime
that has the same ``texture'' when shifted in time, space, or both. How much
one must shift each domain so that its texture maps onto itself is one crude
measure of the structural complexity of the domain.

\begin{Def}
A \emph{domain} is a set of configurations
$\Gamma^{S} = \{ \Gamma_{i,j} \in S \}$, where $S$ is an SR, that satisfies:
\begin{enumerate}
\item Temporal-shift invariance: $\Gamma^{S} = \Phi \Gamma^{S}$, and
\item Spatial-shift invariance: $\Gamma^{S} = \sigma^{\widehat{n}} \Gamma^{S}$
for some spatial offset $\widehat{n}$.
\end{enumerate}
\label{def:Domain}
\end{Def}

It is straightforward to see that, in the spatial population, homogeneous
regions of SRs---$\{T_2\}$ and $\{T_4\}$---are domains. Determining which
additional \eMs---the nonreplaceable ones mentioned above---can be embedded in
these domains, such that the regions are still domains, is a more difficult
calculation that we will not attempt here.

The membranes separating the domains are complementary \eM\ types that actively
function to translate between domains on either side, being composable with
the \eM\ types in those domains. And, over time, sets of membrane \eMs\ map
back into themselves. In the spatial population, as noted above
(Fig. \ref{fig:SpatialMetaMachine}),
$T_1 = T_2 \circ T_4$ and $T_8 = T_4 \circ T_2$. The definition of a membrane
replicator and a membrane express these formally.

\begin{Def}
A set $M$ of \eMs\ is a \emph{membrane replicator} (MR) if 
there are two domain SRs, $S$ and $S^\prime$, such that
\begin{align*}
\theta(S,M,\theta(S,M,S^\prime)) & \subseteq M,\\
\theta(S,M,\theta(S^\prime,M,S)) & \subseteq M,\\
\theta(\theta(S,M,S^\prime),M,S^\prime) & \subseteq M, \mathrm{ and} \\
\theta(\theta(S^\prime,M,S),M,S^\prime) & \subseteq M.
\end{align*}
\label{def:MR}
\end{Def}

Notice that this is similar to the definition of an SR, except that MRs require
the bounding domain SRs to produce \eMs\ in the MR.

\begin{Def}
A \emph{membrane} between two domain SRs $S$ and $S^\prime$ is a
set of configurations:
\begin{equation}
\Gamma^{SS^\prime} = \Gamma^{S} M \Gamma^{S^\prime} ~,
\end{equation}
where $M$ is a membrane replicator, that is temporal-shift invariant:
\begin{equation}
\Gamma^{SS^\prime} = \Phi \Gamma^{SS^\prime} ~.
\end{equation}
\label{def:Membrane}
\end{Def}

Observe that membranes, like domains, are temporally invariant, but, unlike
domains, they are \emph{not} spatially shift invariant. This is what it means,
in fact, for a structure to have a location in space.

Membranes translate between the, possibly distinct, kinds of information stored
in the bounding SR domains. This view parallels that of, for example, the
particles which perform computations in evolved cellular automata
\cite{Crutchfield&Mitchell94a}. Specifically, when the spatial population
dynamic is deterministic, then SR domains and membranes are directly analogous
to the domains and particles of cellular automata computational mechanics
\cite{CACMechMerge}. The particular class of population dynamical
systems considered here differ, however, in important ways: They have an
asynchronous, stochastic spatial update. A more detailed framing of these
differences and the methods to analyze them will be reported elsewhere.

In summary, Defs. \ref{def:Domain}, \ref{def:MR}, and \ref{def:Membrane}
specify algebraic constraints that, in principle, can be solved to find SR
domains and membranes. Even in the simpler case of the computational mechanics
analysis of deterministic CAs, solving the analogous set of constraints,
though well defined, is at present difficult. And so, we will leave detailing
the algorithms for these calculations to a sequel.

\section{Intermediate Spatial Populations}

\begin{figure}[hbtp]
  \includegraphics[width=3.4in]{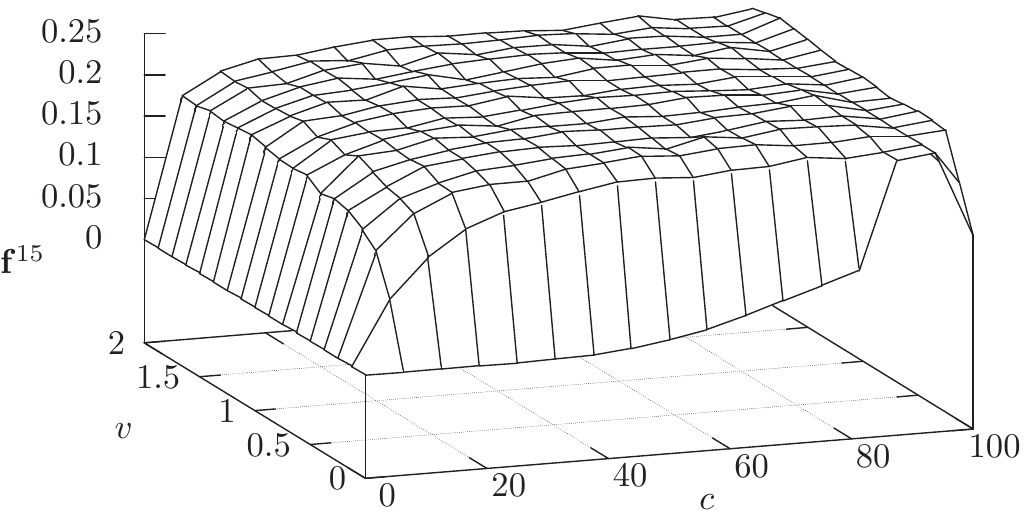}
\caption{$\PrEMType^{15} (t)$ after $t/N = 50$ time steps as a function of
  $c$ and $v$ with $N=10^6$.
  }
\label{fig:IntermediatePlotsSurface}
\end{figure}

The \eM\ soups here were specifically designed to permit investigation of
populations that are neither completely spatial nor completely nonspatial by
varying the mixing parameters $c$ and $v$. To avoid presenting extremely
high-dimensional data, we pick a metric that distinguishes the spatial versus
nonspatial population dynamics and observe how that metric varies with $c$ and
$v$. For the metric we use $\PrEMType^{15} (t)$, the percentage of \eMs\ of type
$T_{15}$ in $\Gamma^t$. In the nonspatial simulation,
$\PrEMType^{15} \approx 0.223$ after $t/N = 50$ steps, while
$\PrEMType^{15} \approx 0.0$ after $t/N = 50$ in the spatial model. In this way,
Fig. \ref{fig:IntermediatePlotsSurface} shows the transition between spatial
and nonspatial behavior for the model as a function of $c$ and $v$. 

\begin{figure}[tp]
  \includegraphics[width=3.4in]{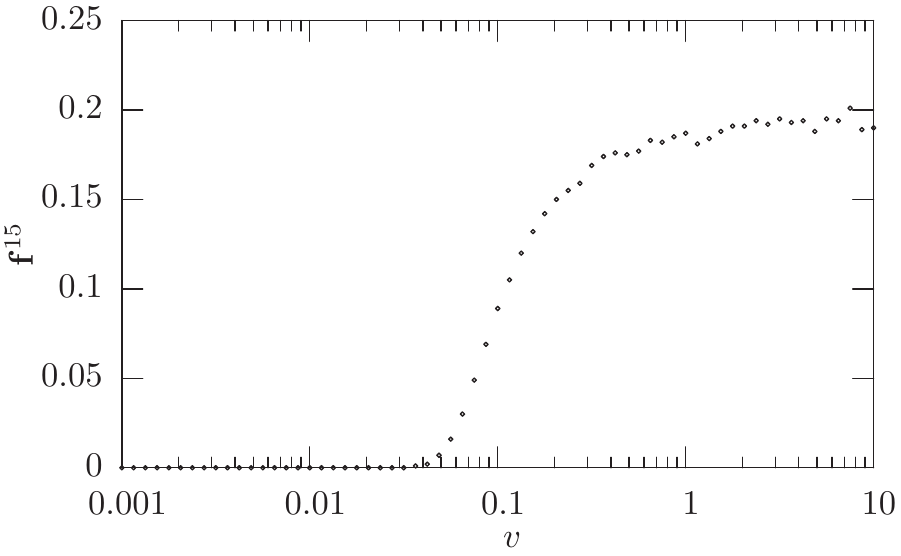}
  \includegraphics[width=3.4in]{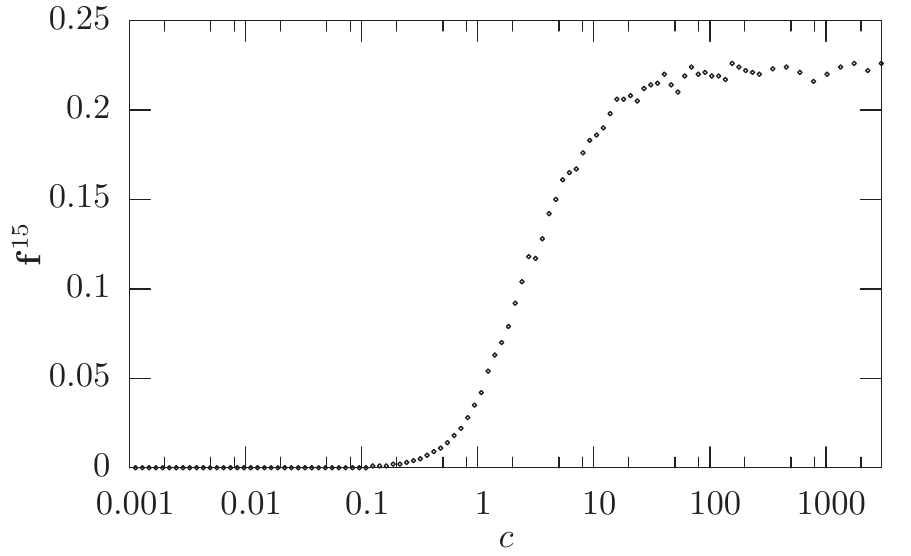}
\caption{Top panel: $\PrEMType^{15} (t)$ after $t/N = 50$ time steps with
  $N=10^6$ and $c=10.0$. Bottom panel: $\PrEMType^{15}$ after $t/N = 50$ time
  steps with $N=10^6$ and $v=1.0$. Note that in both, the horizontal axis is
  logarithmic.
  }
\label{fig:IntermediatePlots}
\end{figure}

The two plots in Fig. \ref{fig:IntermediatePlots} provide a clearer demarcation
of the transition at which spatial memory is lost by showing slices along the
$v$ and $c$ axes. Notably, the plots show that the ``effect of space'' is
logistic in $\log c$ or $\log v$. This indicates that populations are sensitive
to both parameters in similar ways, even though the physical realizations of
varying $c$ (number of \eMs\ swapped) and $v$ (spatial range of mixing) are
quite different. 

\section{Conclusion}

The analysis demonstrated that the capacity for using spatial memory can be
determined by examining the interaction network. Moreover, it also showed
that the behavior of populations which are neither completely spatial nor
completely nonspatial exhibit a systematic dependence on the degree of mixing.
One can imagine several ways to generalize the definition of SRs---as the
definition presented above is sufficient, but by no means necessary---for the
spatial populations to behave differently from the nonspatial populations.

While the exact population dynamics of $\PrEMType$ may be captured by an ODE
such as Eq. (\ref{E:drhoT}), this coarse-graining is not useful to understanding
the population dynamics' key structural properties---the pattern formation
highlighted in the spatial case. Structural analysis of the interaction network,
focusing
specifically on its algebraic properties, was essential to understanding the
mechanisms that drive the spontaneous emergence of organization. To this end,
we drew a useful, if preliminary, connection to the computational mechanics
analysis of domains and walls structures in spatially extended systems.

The membranes observed (and defined) here are not merely concentration
gradients, as seen in familiar pattern formation systems. Rather, they are
entities that, due to their interaction specificity, actively translate between
replicators in neighboring domains. The resulting spatial organization suggests
that one gets spontaneous compartmentalization without designing it in at the
beginning. Recall that compartmentalization is often cited as one of the key
early evolutionary steps on the road to increasing biological complexity
\cite{Mayn95a}. It is not such a difficult step, after all. The \eM\ soup is a
very simple system, built with a minimal set of physical, chemical, and
biological assumptions.

It is perhaps no surprise if we mention that analogous investigations of
one-dimension \eM\ soups leads to similar results. This and the two-dimensional,
intermediate dimensional, and effectively infinite dimensional (panmixia)
soups lead one to wonder about the role of the three-dimensions in which
we know that life arose. Is there something special?

Constructively, in concert with natural spontaneous pattern formation, evolution
may very well commandeer structures of different spatial dimension as ways that
insure various kinds of functionality or intrinsic computation. We appreciate
that evolution is opportunistic and takes advantage of nature's propensity to
spontaneously organize. And so, is three-dimensional space more advantageous
in, say, the range of structures that it supports, compared to one and two
dimensions? Or does evolution take advantage of all possible dimensions?
The empirical evidence speaks rather clearly, if not to the former question,
then to the latter: genetic information is stored in one-dimensional structures,
biological cells sport two-dimensional membranes, and organisms move and
behave in three dimensions.

\section*{Acknowledgments}

SP was supported via the Santa Fe Institute's summer 2004 NSF Research
Experience for Undergraduates Program. This work was partially supported by
CSC's Network Dynamics Program funded by Intel Corporation and by the Defense
Advanced Research Projects Agency (DARPA) Physical Intelligence project. The
views, opinions, and findings contained here are those of the authors and
should not be interpreted as representing the official views or policies,
either expressed or implied, of the DARPA or the Department of Defense.

%

\bibliography{chaos,refs}

\end{document}